\documentclass[seceq]{ptptex}

\title{
\hfill{\normalsize%
\vbox{\hbox{\rm DPNU-03-27}\hbox{\rm December, 2003}  }}\\
\vspace{0.2cm}
Formulation of 
Vector Manifestation in Hot and/or Dense Matter~\footnote{
 Talk given at Finite Density QCD at Nara, 
 July 10-12, 2003, Nara Prefecture Public Hall ``Big Roof",
 Nara, Japan.
}
}

\author{
Masayasu \textsc{Harada}%
}

\inst{
Department of Physics, Nagoya University, Nagoya 464-8602, JAPAN
}

\abst{
The vector manifestation (VM) was proposed as a novel manifestation
of chiral symmetry in which the massless vector meson becomes the
chiral partner of pion.  
In this write-up, I briefly summarize the following
main ingredients to
formulate the VM in hot and/or dense matter:
Effective field theory (EFT) based on the hidden local symmetry;
Wilsonian matching between the EFT and QCD;
Intrinsic thermal and/or dense effects.
}

\begin{document}

\maketitle

\section{Introduction}

Spontaneous chiral symmetry breaking is one of the 
most important properties of QCD in low energy region.
This chiral symmetry is expected to be restored in hot and/or
dense QCD
and properties of hadrons will be changed near the critical
temperature 
of the chiral symmetry restoration~\cite{rest1,Brown-Rho:96,
Rapp-Wambach:00}.
The CERN Super Proton Synchrotron (SPS) observed
an enhancement of dielectron ($e^+e^-$) mass spectra
below the $\rho / \omega$ resonance~\cite{Agakishiev:1995xb}.
This can be explained by the dropping mass of the $\rho$ meson
(see, e.g., Refs.~\citen{Li:1995qm, Brown-Rho:96, Rapp-Wambach:00})
following the Brown-Rho scaling proposed in Ref.~\citen{BR}.
Furthermore,
the Relativistic Heavy Ion Collider (RHIC) has started
to measure several physical processes in hot matter
which include the dilepton energy spectra.
Therefore it is interesting to study the temperature and/or
density dependences
of the vector meson mass which is one of the important quantities
in the chiral phase transition.

In Ref.~\citen{HS:VMT},
it was shown how the vector manifestation (VM)~\cite{HY:VM}, 
in which the chiral symmetry is restored by 
the massless degenerate pseudoscalar meson denoted by $\pi$
(pion and its flavor partners) and the
vector meson denoted by $\rho$ 
($\rho$ meson and its flavor partners) as the chiral partner,
is formulated in hot matter using the model for $\pi$ and
$\rho$ based on the
hidden local symmetry (HLS)~\cite{BKUYY}.
Furthermore, in Ref.~\citen{HKR}, the formulation
of the VM is done in the presence of dense matter 
by including the quasiquark into the HLS.
In these formulations, 
the {\it intrinsic temperature and/or density
dependences}~\cite{HS:VMT,HKR}
of the parameters of the HLS Lagrangian,
which is introduced by applying
the Wilsonian matching~\cite{HY:WM,HY:PRep} at 
non-zero temperature and/or non-zero density,
played important roles
to realize the chiral symmetry restoration consistently:
In the framework of the HLS the equality between 
the axial-vector and vector current correlators at critical point
can be satisfied only if the intrinsic thermal and/or density
effects are included.

In this write-up, I show how the VM is formulated 
in hot and/or dense QCD following Refs.~\citen{HS:VMT,HS:VD,HKR}.
I first show the difference between the VM and
the conventional picture based on the linear sigma model
in terms of the chiral representation of the low-lying mesons
in section~\ref{sec:VMCS}.
I will introduce the model based on 
the HLS, and show the renormalization group equations for the
parameters of the HLS Lagrangian
in section~\ref{sec:EFT}.  
I will also briefly summarize
the general idea of 
Wilsonian matching between the effective field theory
and QCD.
In section~\ref{sec:ITDE},
the intrinsic thermal and/or density effects are briefly explained.
Formulations of the VM in hot and dense matter are shown in 
sections~\ref{sec:VMT} and \ref{sec:VMD}.
Finally, in section~\ref{sec:SD}, I will give a brief summary
and discussions.

\section{Vector Manifestation of Chiral Symmetry}
\label{sec:VMCS}

In this section, following Ref.~\citen{HY:VM,HY:PRep},
I briefly review the difference between the 
vector manifestation (VM) and
the conventional manifestation of chiral symmetry restoration
based on the linear sigma model
in terms of the chiral representation of the mesons
by extending the analyses done in
Refs.~\citen{Gilman-Harari,Weinberg:69}
for two flavor QCD.

The VM was first proposed in
Ref.~\citen{HY:VM} as a novel manifestation of Wigner 
realization of
chiral symmetry where the vector meson $\rho$ becomes massless at the
chiral phase transition point. 
Accordingly, the (longitudinal) $\rho$ becomes the chiral partner of
the Nambu-Goldstone (NG) boson $\pi$.
The VM is characterized by
\begin{equation}
\mbox{(VM)} \qquad
f_\pi^2 \rightarrow 0 \ , \quad
m_\rho^2 \rightarrow m_\pi^2 = 0 \ , \quad
f_\rho^2 / f_\pi^2 \rightarrow 1 \ ,
\label{VM def}
\end{equation}
where $f_\rho$ is the decay constant of 
(longitudinal) $\rho$ at $\rho$ on-shell.
This is completely different from 
the conventional picture based
on the linear sigma model 
where the scalar meson $S$ becomes massless
degenerate with $\pi$ as the chiral partner:
\begin{equation}
\mbox{(GL)} \qquad
f_\pi^2 \rightarrow 0 \ , \quad
m_S^2 \rightarrow m_\pi^2 = 0 \ .
\label{GL def}
\end{equation}
In Ref.~\citen{HY:PRep}
this was called GL manifestation after the
effective theory of Ginzburg--Landau or Gell-Mann--Levy.

I first consider 
the representations of 
the following zero helicity ($\lambda=0$) states
under
$\mbox{SU(3)}_{\rm L}\times\mbox{SU(3)}_{\rm R}$;
the $\pi$, the (longitudinal) $\rho$, the (longitudinal) axial-vector
meson denoted by $A_1$ ($a_1$ meson and its flavor partners)
and the scalar meson denoted by $S$.
The $\pi$ and the longitudinal $A_1$ 
are admixture of $(8\,,\,1) \oplus(1\,,\,8)$ and 
$(3\,,\,3^*)\oplus(3^*\,,\,3)$
since the symmetry is spontaneously
broken~\cite{Weinberg:69,Gilman-Harari}:
\begin{eqnarray}
\vert \pi\rangle &=&
\vert (3\,,\,3^*)\oplus (3^*\,,\,3) \rangle \sin\psi
+
\vert(8\,,\,1)\oplus (1\,,\,8)\rangle  \cos\psi
\ ,
\nonumber
\\
\vert A_1(\lambda=0)\rangle &=&
\vert (3\,,\,3^*)\oplus (3^*\,,\,3) \rangle \cos\psi 
- \vert(8\,,\,1)\oplus (1\,,\,8)\rangle  \sin\psi
\ ,
\label{mix pi A}
\end{eqnarray}
where the experimental value of the mixing angle $\psi$ is 
given by approximately 
$\psi=\pi/4$~\cite{Weinberg:69,Gilman-Harari}.  
On the other hand, the longitudinal $\rho$
belongs to pure $(8\,,\,1)\oplus (1\,,\,8)$
and the scalar meson to 
pure $(3\,,\,3^*)\oplus (3^*\,,\,3)$:
\begin{eqnarray}
\vert \rho(\lambda=0)\rangle &=&
\vert(8\,,\,1)\oplus (1\,,\,8)\rangle  
\ ,
\nonumber
\\
\vert S\rangle &=&
\vert (3\,,\,3^*)\oplus (3^*\,,\,3) \rangle 
\ .
\label{rhoS}
\end{eqnarray}

When the chiral symmetry is restored at the
phase transition point, 
it is natural to expect that
the chiral representations coincide with the mass eigenstates:
The representation mixing is dissolved.
{}From Eq.~(\ref{mix pi A}) one can easily see~\cite{HY:VM}
that
there are two ways to express the representations in the
Wigner phase of the chiral symmetry:
The conventional GL manifestation
corresponds to 
the limit $\psi \rightarrow \pi/2$ in which
$\pi$ is in the representation
of pure $(3\,,\,3^*)\oplus(3^*\,,\,3)$ 
together with the scalar meson, 
both being the chiral partners:
\begin{eqnarray}
\mbox{(GL)}
\qquad
\left\{
\begin{array}{rcl}
\vert \pi\rangle\,, \vert S\rangle
 &\rightarrow& 
\vert  (3\,,\,3^\ast)\oplus(3^\ast\,,\,3)\rangle\ ,
\\
\vert \rho (\lambda=0) \rangle \,,
\vert A_1(\lambda=0)\rangle  &\rightarrow&
\vert(8\,,\,1) \oplus (1\,,\,8)\rangle\ .
\end{array}\right.
\end{eqnarray}
On the other hand, the VM corresponds 
to the limit $\psi\rightarrow 0$ in which the $A_1$ 
goes to a pure 
$(3\,,\,3^*)\oplus (3^*\,,\,3)$, now degenerate with
the scalar meson $S$ in the same representation, 
but not with $\rho$ in 
$(8\,,\,1)\oplus (1\,,\,8)$:
\begin{eqnarray}
\mbox{(VM)}
\qquad
\left\{
\begin{array}{rcl}
\vert \pi\rangle\,, \vert \rho (\lambda=0) \rangle
 &\rightarrow& 
\vert(8\,,\,1) \oplus (1\,,\,8)\rangle\ ,
\\
\vert A_1(\lambda=0)\rangle\,, \vert S\rangle  &\rightarrow&
\vert  (3\,,\,3^\ast)\oplus(3^\ast\,,\,3)\rangle\ .
\end{array}\right.
\end{eqnarray}
Namely, the
degenerate massless $\pi$ and (longitudinal) $\rho$ at the 
phase transition point are
the chiral partners in the
representation of $(8\,,\,1)\oplus (1\,,\,8)$.

Next, I consider the helicity $\lambda=\pm1$. 
Note that
the transverse $\rho$
can belong to the representation different from the one
for the longitudinal $\rho$ ($\lambda=0$) and thus can have the
different chiral partners.
According to the analysis in Ref.~\citen{Gilman-Harari},
the transverse components of $\rho$ ($\lambda=\pm1$)
in the broken phase
belong to almost pure
$(3^*\,,\,3)$ ($\lambda=+1$) and $(3\,,\,3^*)$ ($\lambda=-1$)
with tiny mixing with
$(8\,,\,1)\oplus(1\,,\,8)$.
Then, it is natural to consider in VM that
they become pure $(3\,,\,3^\ast)$ and 
$(3^\ast\,,\,3)$
in the limit approaching the chiral restoration point~\cite{HY:PRep}:
\begin{eqnarray}
\vert \rho(\lambda=+1)\rangle \rightarrow 
  \vert (3^*,3)\rangle\ ,\quad
\vert \rho(\lambda=-1)\rangle \rightarrow 
  \vert (3,3^*)\rangle \ .
\end{eqnarray}
As a result,
the chiral partners of the transverse components of $\rho$ 
in the VM
will be  themselves. Near the critical point the longitudinal $\rho$
becomes 
almost $\sigma$, namely the would-be NG boson $\sigma$ almost 
becomes a 
true NG boson and hence a different particle than the transverse
$\rho$.

\section{Effective Field Theory}
\label{sec:EFT}

In this section I first show the effective field theory (EFT)
in which
the vector manifestation is formulated (subsection~\ref{ssec:HLS}).
Then, after showing the renormalization group equations (RGEs) for
the parameters of the Lagrangian of the EFT
(subsection~\ref{ssec:RGE}),
I briefly summarize a 
general idea 
of the Wilsonian matching between the EFT 
and QCD which determines the 
parameters of the Lagrangian of the EFT (subsection~\ref{ssec:WM}).

I should note that,
as is stressed in Ref.~\citen{HY:PRep},
the VM can be formulated only as a limit by approaching it from the
broken phase of chiral symmetry.
Then, for the formulation of the VM,
I need an EFT
including $\rho$ and $\pi$ 
in the broken phase which is not
necessarily applicable in the symmetric phase.
One of such EFTs is the model based on the 
hidden local symmetry (HLS)~\cite{BKUYY}
which includes $\rho$ as the gauge boson of the HLS in addition
to $\pi$ as the Nambu-Goldstone (NG) boson associated with the
chiral symmetry breaking in a manner fully consistent with
the chiral symmetry of QCD.
It should be noticed that,
in the HLS,
thanks to the gauge invariance
one can perform the
systematic chiral perturbation with including $\rho$
in addition to $\pi$.~\cite{Georgi,HY:PLB,Tanabashi,HY:WM,HY:PRep}

\subsection{Hidden Local Symmetry}
\label{ssec:HLS}

The HLS model is based on 
the $G_{\rm{global}} \times H_{\rm{local}}$ symmetry,
where $G=\mbox{SU}(N_f)_{\rm L} \times \mbox{SU}(N_f)_{\rm R}$ 
is the chiral symmetry
and $H=\mbox{SU}(N_f)_{\rm V}$ is the HLS. 
The basic quantities are 
the HLS gauge boson $V_\mu$ and two matrix valued
variables $\xi_{\rm L}(x)$ and $\xi_{\rm R}(x)$
which transform as
 \begin{equation}
  \xi_{\rm L,R}(x) \to \xi^{\prime}_{\rm L,R}(x)
  =h(x)\xi_{\rm L,R}(x)g^{\dagger}_{\rm L,R}\ ,
 \end{equation}
where $h(x)\in H_{\rm{local}}\ \mbox{and}\ g_{\rm L,R}\in
[\mbox{SU}(N_f)_{\rm L,R}]_{\rm{global}}$.
These variables are parameterized as
 \begin{equation}
  \xi_{\rm L,R}(x)=
    e^{i\sigma (x)/{F_\sigma}}e^{\mp i\pi (x)/{F_\pi}}\ ,
 \end{equation}
where $\pi = \pi^a T_a$ denotes the pseudoscalar 
NG bosons
associated with the spontaneous symmetry breaking of
$G_{\rm{global}}$ chiral symmetry, 
and $\sigma = \sigma^a T_a$ denotes
the NG bosons associated with 
the spontaneous breaking of $H_{\rm{local}}$.
This $\sigma$ is absorbed into the HLS gauge 
boson through the Higgs mechanism. 
$F_\pi \ \mbox{and}\ F_\sigma$ are the decay constants
of the associated particles.
The phenomenologically important parameter $a$ is defined as 
 \begin{equation}
  a = \frac{{F_\sigma}^2}{{F_\pi}^2}\ .
 \end{equation}
The covariant derivatives of $\xi_{\rm L,R}$ are given by
\begin{eqnarray}
 D_\mu \xi_{\rm L} &=& 
  \partial_\mu\xi_{\rm L} - iV_\mu \xi_{\rm L} 
  + i\xi_{\rm L} {\cal{L}}_\mu\ ,
 \nonumber\\
 D_\mu \xi_{\rm R} &=& 
  \partial_\mu\xi_{\rm R} - iV_\mu \xi_{\rm R} 
  + i\xi_{\rm R} {\cal{R}}_\mu \ ,
\end{eqnarray}
where $V_\mu$ is the gauge field of $H_{\rm{local}}$, and
${\cal{L}}_\mu \ \mbox{and}\ {\cal{R}}_\mu$ are the external
gauge fields introduced by gauging $G_{\rm{global}}$ symmetry.

The HLS Lagrangian with lowest derivative terms 
is given by~\cite{BKUYY}
 \begin{equation}
  {\cal{L}}_{(2)} = {F_\pi}^2\mbox{tr}\bigl[ \hat{\alpha}_{\perp\mu}
                                      \hat{\alpha}_{\perp}^{\mu}
                                   \bigr] +
       {F_\sigma}^2\mbox{tr}\bigl[ \hat{\alpha}_{\parallel\mu}
                  \hat{\alpha}_{\parallel}^{\mu}
                  \bigr] -
        \frac{1}{2g^2}\mbox{tr}\bigl[ V_{\mu\nu}V^{\mu\nu}
                   \bigr]
\ , \label{eq:L(2)}
 \end{equation}
where $g$ is the HLS gauge coupling,
$V_{\mu\nu}$ is the field strength
of $V_\mu$ and
\begin{eqnarray}
&&
\hat{\alpha}_{\perp,\parallel}^\mu =
( D_\mu \xi_{\rm R} \cdot \xi_{\rm R}^\dag \mp 
  D_\mu \xi_{\rm L} \cdot \xi_{\rm L}^\dag
) / (2i)
\ .
\end{eqnarray}

\subsection{Renormalization Group Equations}
\label{ssec:RGE}

At one-loop level
the Lagrangian (\ref{eq:L(2)}) 
generates the ${\mathcal O}(p^4)$
contributions including the divergent contributions which
are renormalized by three leading order
parameters $F_\pi$, $a$ and $g$ (and parameters of 
${\mathcal O}(p^4)$ Lagrangian).
As was stressed in Ref.~\citen{HY:PRep},
it is important to include effects of quadratic divergences
into the RGEs
for studying the phase structure.
The resultant RGEs for three leading order parameters
are expressed as~\cite{HY:conformal,HY:WM,HY:PRep}
\begin{eqnarray}
{\mathcal M} \frac{dF_\pi^2}{d{\mathcal M}} &=& 
C\left[3a^2g^2F_\pi^2 +2(2-a){\mathcal M}^2 \right] 
\ ,
\nonumber\\
{\mathcal M} \frac{da}{d{\mathcal M}} &=&-C
(a-1) \left [3a(1+a)g^2-(3a-1)\frac{{\mathcal M}^2}{F_\pi^2} \right]
\ ,
\nonumber\\
{\mathcal M}\frac{d g^2}{d {\mathcal M}}&=& -C\frac{87-a^2}{6}g^4
\ ,
\label{rge}
\end{eqnarray}
where $C = N_f/\left[2(4\pi)^2\right]$ and ${\mathcal M}$ is the
renormalization point.
It should be noted that the point $(g\,,\,a)=(0,1)$
is the fixed point of 
the RGEs in Eq.~(\ref{rge})
which plays an essential role to formulate the VM in the following
analysis of the chiral symmetry restoration.

\subsection{Wilsonian Matching}
\label{ssec:WM}

The basic concept of the EFT is that the effective Lagrangian,
which has the most general form constructed from the chiral symmetry,
give the same generating functional as that obtained from QCD:
\begin{equation}
 Z_{\rm EFT}[J,F] = \int {\cal D}U e^{i S_{\rm eff}[J,F]}
 \mathop{\Leftrightarrow}_{\rm matching}
 Z_{\rm QCD}[J] = \int {\cal D}q {\cal D}\bar{q} {\cal D}G
         e^{i S_{\rm QCD}[J]}
\ ,
\label{gf-match}
\end{equation}
where $J$ is a set of external source fields.
In the EFT side
$U$ denotes the relevant hadronic fields such as the pion fields,
$S_{\rm eff}$ is the action 
expressed in terms of these hadrons,
$F$ a set of parameters included in the EFT.
In QCD side
$q$ $(\bar{q})$ denotes (anti) quark field,
$G$ is gluon field and $S_{\rm QCD}$ represents the action
expressed in terms of the quarks and gluons.

In some matching schemes, the renormalized quantities of the EFT
are determined from QCD.
On the other hand,
the matching in the Wilsonian sense is performed based on the 
following general idea:
The bare Lagrangian of the EFT is defined at a suitable 
matching scale $\Lambda$ and
the generating functional derived from the bare Lagrangian 
leads to the same Green's function as that derived in QCD
at $\Lambda$:
\begin{equation}
\bigl.
 Z_{\rm EFT}[J,F] 
\bigr\vert_{E = \Lambda}
= e^{i S_{\rm eff}[J,F_{\rm bare}]}
 \mathop{\Leftrightarrow}_{\rm matching}
\bigl.
 Z_{\rm QCD}[J] 
\bigr\vert_{E = \Lambda}
= \int {\cal D}q {\cal D}\bar{q} {\cal D}G
         e^{i S_{\rm QCD}[J]}
\ ,
\label{gf-bare-match}
\end{equation}
where $F_{\rm bare}$ denotes a set of bare parameters.
Through the above matching, which is named {\it Wilsonian matching}
in Ref.~\citen{HY:WM},
the {\it bare} parameters of the EFT are determined.
In other words,
we obtain the bare Lagrangian of the EFT after
integrating out the high energy modes, i.e.,
the quarks and gluons above $\Lambda$.
Then the informations of the high energy modes are included in the
parameters of the EFT.

In Ref.~\citen{HY:WM,HY:PRep}, based on the above idea, 
the vector and axial-vector current correlators derived
from the bare HLS theory are matched with those obtained
by the operator product expansion in QCD.
It was shown that the physical predictions are in remarkable
agreement with experiments.

\section{Intrinsic Thermal and/or Density Effects}
\label{sec:ITDE}

In Refs.~\citen{HS:VMT,HKR} 
the 
Wilsonian matching, briefly explained in subsection~\ref{ssec:WM},
was applied to the analysis of QCD in hot and dense matter.
As was discussed in subsection~\ref{ssec:WM}, 
the bare Lagrangian of the effective field theory
(EFT) is obtained by integrating out high
energy modes, i.e., quarks and gluons above the matching scale
$\Lambda$, and as as result,
the informations of the high energy modes are
included into the
parameters of the EFT.
Thus when we integrate out high energy modes in hot and/or dense
matter,
the parameters are in general dependent on temperature and/or density.
This is called the {\it intrinsic temperature and/or density
dependence}~\cite{HS:VMT,HKR}, which is
nothing but the signature that hadron has an internal structure 
constructed from the quarks and gluons.
This is similar to the situation where the coupling constants
among hadrons are replaced with the momentum-dependent form factor
in high energy region.
Thus the intrinsic thermal and/or dense effects play more important
roles in higher temperature region, 
especially near the critical temperature.

Here, let me show an example of the Wilsonian matching condition
to determine the bare decay constant of $\pi$
in hot and/or dense matter.
This is obtained~\cite{HS:VMT,HKR} by
putting possible temperature and/or density
dependences on the gluonic and quark condensates in 
the Wilsonian matching condition at 
$T=\mu=0$~\footnote{
 It should be noticed that there is no longer Lorentz symmetry
 in hot and/or dense matter, and
 the Lorentz non-scalar operators such as
 $\bar{q}\gamma_\mu D_\nu q$ may exist in 
 the form of the current correlators derived by the 
 operator product expansion~\cite{HKL}.
 This leads to, e.g., a difference between the temporal and spatial
 bare pion decay constants.
 However, I neglect the contributions from these operators here
 since they give a small correction compared with 
 the main term $1 + \frac{\alpha_s}{\pi}$.
 This implies that the Lorentz symmetry breaking effect in
 the bare pion decay constant is small, 
 $F_{\pi,\rm{bare}}^t \simeq F_{\pi,\rm{bare}}^s$~\cite{HKRS}.
 The Wilsonian matching with including the effect of such
 Lorentz non-scalar operators was recently done in 
 Ref.~\citen{HKRS:pvT}, which shows that the difference between
 $F_{\pi,\rm{bare}}^t$ and 
 $F_{\pi,\rm{bare}}^s$ is actually small.
}:
\begin{equation}
 \frac{F^2_\pi (\Lambda ;T,\mu)}{{\Lambda}^2} 
  = \frac{1}{8{\pi}^2}\Bigl[ 1 + \frac{\alpha _s}{\pi} +
     \frac{2{\pi}^2}{3}\frac{\langle \frac{\alpha _s}{\pi}
      G_{\mu \nu}G^{\mu \nu} \rangle_{T,\mu} }{{\Lambda}^4} +
     {\pi}^3 \frac{1408}{27}\frac{\alpha _s{\langle \bar{q}q
      \rangle }^2_{T,\mu}}{{\Lambda}^6} \Bigr]
\ .
\label{eq:WMC A}
\end{equation}
Through this condition
the temperature and/or density
dependences of the quark and gluonic condensates
determine the intrinsic temperature dependences 
of the bare parameter $F_\pi(\Lambda;T,\mu)$,
which is then converted into 
those of the on-shell parameter $F_\pi(\mu=0;T,\mu)$ 
through the Wilsonian renormalization group equations.

\section{Vector Manifestation in Hot Matter}
\label{sec:VMT}

In this section, I show how the vector manifestation (VM) is
formulated in hot QCD following Refs.~\citen{HS:VMT,HS:VD}.

Let me start from 
the axial-vector and the vector
current correlators derived from the bare hidden local symmetry
(HLS):
\begin{eqnarray}
 G^{\rm{(HLS)}}_A (Q^2) 
  &=& \frac{F^2_\pi (\Lambda;T)}{Q^2} -
      2z_2(\Lambda;T), \nonumber\\
 G^{\rm{(HLS)}}_V (Q^2) 
  &=& \frac{F^2_\sigma (\Lambda;T)
         [1 - 2g^2(\Lambda;T)z_3(\Lambda;T)]}
           {{M_\rho}^2(\Lambda;T) + Q^2} - 2z_1(\Lambda;T)\ .
  \label{correlator HLS at zero-T}
  \end{eqnarray}
It should be noticed that 
the the bare parameters have the 
intrinsic temperature dependences~\cite{HS:VMT}
as I explained in the previous section.

At the critical temperature,
the axial-vector and vector current correlators
derived in the operator product expansion (OPE)
agree with each other for any value of $Q^2$.
Thus I require that
these current correlators in the HLS are
equal at the critical temperature
for any value of $Q^2\ \mbox{around}\ {\Lambda}^2$.
By taking account of the fact 
$F^2_\pi (\Lambda ;T_c) \neq 0$ derived by applying
the Wilsonian matching condition in Eq.~(\ref{eq:WMC A})
at $T=T_c$,
the requirement 
$G_A^{(\rm{HLS})}=G_V^{(\rm{HLS})}$ is satisfied
only if the following conditions are met~\cite{HS:VMT}: 
\begin{eqnarray}
&&
g(\Lambda;T) \mathop{\longrightarrow}_{T \rightarrow T_c} 0 \ ,
\qquad
a(\Lambda;T) \mathop{\longrightarrow}_{T \rightarrow T_c} 1 \ ,
\qquad
z_1(\Lambda;T) - z_2(\Lambda;T) 
\mathop{\longrightarrow}_{T \rightarrow T_c} 0 \ .
\label{g a z12:VMT}
\end{eqnarray}
These conditions (``VM conditions in hot matter'')
for the bare parameters
are converted into the
conditions for the on-shell parameters through the Wilsonian 
renormalization group equations (RGEs).
Since $g=0$ and $a=1$ are separately the fixed points of the RGEs for
$g$ and $a$~\cite{HY:conformal},
the on-shell parameters also satisfy
$(g,a)=(0,1)$, and thus the parametric $\rho$ mass
satisfies $M_\rho = 0$.

Now, let me
include the hadronic thermal effects to obtain the $\rho$ pole
mass near the critical temperature.
As I explained above,
the intrinsic temperature dependences imply that
$M_\rho/T \rightarrow 0$
for $T \rightarrow T_c$,
so that the $\rho$ pole mass near the
critical temperature is expressed as~\cite{HS:VMT,HS:VD}
\begin{eqnarray}
&& m_\rho^2(T)
  = M_\rho^2 +
  g^2 N_f \, \frac{15 - a^2}{144} \,T^2
\ .
\label{mrho at T 2}
\end{eqnarray}
Since $a \simeq 1$ near the restoration point,
the second term is positive. 
Then the $\rho$ pole mass $m_\rho$
is bigger than the parametric
$M_\rho$ due to the hadronic thermal corrections.
Nevertheless, 
{\it the intrinsic temperature dependence determined by the
Wilsonian matching requires
that the $\rho$ becomes massless at the
critical temperature}:
\begin{eqnarray}
&&
m_\rho^2(T)
\rightarrow 0 \ \ \mbox{for} \ T \rightarrow T_c \ ,
\end{eqnarray}
since the first term in Eq.~(\ref{mrho at T 2})
vanishes as $M_\rho\rightarrow 0$, and the second
term also vanishes since $g\rightarrow 0$ for $T \rightarrow T_c$.
This implies that
{\it the vector manifestation (VM) actually
occurs at the critical
temperature}~\cite{HS:VMT}.

\section{Vector Manifestation in Dense Matter}
\label{sec:VMD}

In this section, I briefly summarize how the vector manifestation
(VM) is formulated in dense QCD following Ref.~\citen{HKR}.

In Ref.~\citen{HKR}, 
following the picture shown in Ref.~\citen{Brown-Rho:01b},
the quasiquark degree of freedom is added into the Lagrangian
of the hidden local symmetry (HLS)
near the critical chemical potential
with assuming that its mass $m_q$ becomes small 
($m_q \rightarrow 0$).
The Lagrangian introduced in Ref.~\citen{HKR} for
including one quasiquark field $\psi$
and one anti-quasiquark field $\bar{\psi}$ is
counted as ${\mathcal O}(p)$ and given by
\begin{eqnarray}
 {\mathcal L}_{Q} 
= 
 \bar \psi( iD_\mu \gamma^\mu
       + \mu \gamma^0 -m_q )\psi
  {}+ \bar \psi \left(
  \kappa\gamma^\mu \hat{\alpha}_{\parallel \mu}
+ \lambda\gamma_5\gamma^\mu \hat{\alpha}_{\perp\mu} \right)
        \psi 
\ ,
\label{lagbaryon}
 \end{eqnarray}
where $\mu$ is the chemical potential,
$D_\mu\psi=(\partial_\mu -i g \rho_\mu)\psi$ and $\kappa$ and
$\lambda$ are constants to be specified later.

Inclusion of the quasiquark changes the renormalization
group equations (RGEs) for $F_\pi$, $a$ 
and $g$.
Furthermore, the RGE for the quasiquark mass $m_q$ should be
considered simultaneously.
The explicit forms of the RGEs are shown in Eq.~(7)
of Ref.~\citen{HKR},
which show 
that,
although $g=0$ and $a=1$ are not separately
the fixed points of the RGEs for $g$ and $a$, 
$(g,a,m_q) = (0,1,0)$ is a fixed point of the coupled RGEs for 
$g$, $a$ and $m_q$.

Let me consider the 
{\it intrinsic density dependences} of the 
bare parameters of the HLS Lagrangian.
Similarly to the intrinsic temperature dependences in hot QCD,
the intrinsic density dependences are 
introduced through the Wilsonian matching.
Noting that the quasiquark does not contribute to the current
correlators at {\it bare level}, one arrives at the following
``VM conditions in dense matter'' similar to the one
in hot matter in Eq.~(\ref{g a z12:VMT}) near the critical 
chemical potential $\mu_c$~\cite{HKR}:
\begin{eqnarray}
&&
g(\Lambda;\mu) \mathop{\longrightarrow}_{\mu \rightarrow \mu_c} 0 \ ,
\qquad
a(\Lambda;\mu) \mathop{\longrightarrow}_{\mu \rightarrow \mu_c} 1 \ ,
\qquad
z_1(\Lambda;\mu) - z_2(\Lambda;\mu) 
\mathop{\longrightarrow}_{\mu \rightarrow \mu_c} 0 \ .
\label{g a z12:VMm}
\end{eqnarray}
These conditions are converted into the conditions for the
on-shell parameters through the RGEs.
Since 
$(g,a,m_q) = (0,1,0)$ is a fixed point of the coupled RGEs for 
$g$, $a$ and $m_q$, the above conditions together with the assumption
$m_q \rightarrow 0$ for $\mu \rightarrow \mu_c$ imply that
the on-shell parameters behave as
$g\rightarrow 0$ and $a\rightarrow 1$, and thus the parametric
$\rho$ mass vanishes for $\mu \rightarrow \mu_c$: 
$M_\rho \rightarrow 0$.

Now, let me study the $\rho$ pole mass near $\mu_c$
by including the hadronic dense-loop correction from the
quasiquark.
By taking $(g,a,m_q) \rightarrow (0,1,0)$ 
in the quasiquark loop contribution, 
the $\rho$ pole mass is expressed as
\begin{equation}
m_\rho^2(\mu) = M_\rho^2(\mu) + g^2 \,
\frac{(1-\kappa)^2}{6\pi^2} \,\mu^2 \ .
\end{equation}
Since $M_\rho(\mu) \rightarrow 0$ and $g\rightarrow0$ for
$\mu \rightarrow \mu_c$ due to the intrinsic density dependence,
the above expression implies that 
the $\rho$ pole mass vanishes at the critical chemical potential,
i.e.,
the VM is realized in dense matter:
\begin{equation}
m_\rho(\mu) 
\rightarrow 0 \ \ \mbox{for} \ \mu \rightarrow \mu_c \ .
\end{equation}

\section{Summary and Discussions}
\label{sec:SD}

In this write-up I summarized how the vector manifestation
(VM) is formulated in hot and/or dense matter based on the
hidden local symmetry (HLS).
One of the most important ingredients to formulate the VM 
in hot and/or dense matter is
the {\it intrinsic temperature and/or density 
dependence}~\cite{HS:VMT,HKR} of the {\it bare} parameters
of the bare HLS Lagrangian derived through the Wilsonian
matching~\cite{HY:WM,HY:PRep}.
For satisfying the
equality between the axial-vector and vector current correlators
at the chiral restoration point ($T=T_c$ and/or $\mu=\mu_c$),
which is needed for consistency with the chiral symmetry
restoration, 
the intrinsic temperature and/or density dependence leads to
the VM conditions
in Eq.~(\ref{g a z12:VMT}) and/or Eq.~(\ref{g a z12:VMm}).
These conditions are protected by the VM fixed point
of the
renormalization group equations, and play crucial role to 
formulate the VM in hot and/or dense matter.

In this write-up, I explained only the formulation of the VM,
and did not introduce the following 
several predictions of the VM in hot matter
done so far:
The vector and axial-vector susceptibilities are predicted to be
equal~\cite{HKRS};
The pion velocity becomes the speed of light when we neglect 
the small Lorentz violating effects in the bare HLS 
Lagrangian~\cite{HKRS};
the vector dominance
of the electromagnetic form factor of pion
is largely violated~\cite{HS:VD}.
Recently in Ref.~\citen{Sasaki:Vpi}, 
Sasaki included the Lorentz breaking
effects into the bare HLS Lagrangian, and showed
that the pion velocity at the critical temperature receives neither
quantum nor hadronic thermal corrections.
In Ref.~\citen{HKRS:pvT},
based on this ``non-renormalization theorem'' and the
Lorentz non-invariant version of the Wilsonian matching,
the pion velocity near the critical temperature was shown 
to be close to the speed of light.
This
result is drastically different from the result obtained in
the standard chiral theory~\cite{SS}
for the
two-flavor QCD which predicts that the pion velocity should go to
zero at the critical point.

\section*{Acknowledgements}
I would like to thank the organizers of this workshop 
for giving me an opportunity to present my talk.
I am grateful to Dr. Youngman Kim, Professor Mannque Rho,
Dr. Chihiro Sasaki and Professor Koichi Yamawaki for collaboration
in the works on which this write-up is based.
This work is 
supported in part by the 21st Century COE
Program of Nagoya University provided by Japan Society for the
Promotion of Science (15COEG01).

%

\end{document}